\documentclass{article}
\usepackage{graphicx} 
\usepackage[table]{xcolor}
\usepackage{color, colortbl}

\usepackage{authblk}
\providecommand{\keywords}[1]{\textbf{\textit{Keywords:}} #1}

\usepackage{geometry}
\usepackage{booktabs}
\usepackage{caption}
\usepackage{algorithm}
\usepackage{algpseudocode}

\title{Dynamic Data Defense: Unveiling the Database in motion Chaos Encryption (DaChE) Algorithm — A Breakthrough in Chaos Theory for Enhanced Database Security}

\author[1]{Abraham Itzhak Weinberg}
\affil[1]{AI-WEINBERG, AI Experts, Tel Aviv, Israel, aviw2010@gmail.com}

\begin{document}
\maketitle
\begin{abstract}
Amidst the burgeoning landscape of database architectures, the surge in NoSQL databases has heralded a transformative era, liberating data storage from traditional relational constraints and ushering in unprecedented scalability. As organizations grapple with the escalating security threats posed by database breaches, a novel theoretical framework emerges at the nexus of chaos theory and topology: the Database in motion Chaos Encryption (DaChE) Algorithm. This paradigm-shifting approach challenges the static nature of data storage, advocating for dynamic data motion to fortify database security. By incorporating chaos theory, this innovative strategy not only enhances database defenses against evolving attack vectors but also redefines the boundaries of data protection, offering a paradigmatic shift in safeguarding critical information assets. Additionally, it enables parallel processing, facilitating on-the-fly processing and optimizing the performance of the proposed framework.
\end{abstract}


\section{Introduction}
In today's data-driven world, centralized database systems form the backbone of organizational infrastructure, storing valuable assets that drive business decisions and operations. The rise of NoSQL databases and big data architectures has transformed how organizations store and process information. However, these advances have introduced new security challenges. In 2024, the average cost of a data breach reached a record high of \$4.88 million, reflecting a 10\% increase from the previous year \cite{secureframe2024}. This alarming trend underscores the pressing need for innovative database security solutions.\\
Traditional security methods, such as conventional encryption, have proven insufficient in addressing evolving threats. While encryption is essential for protecting sensitive data, it introduces performance overhead and is vulnerable to emerging attack vectors. Despite advances in quantum-proof encryption, studies have identified vulnerabilities even in these sophisticated cryptographic techniques \cite{schneier2024lattice}. The challenge lies in the static nature of these security measures, which fail to adapt dynamically to new and real-time threats.\\
Furthermore, modern database systems, particularly NoSQL architectures, prioritize scalability and flexibility over traditional security measures. Although they excel at handling massive volumes of unstructured data, they often sacrifice robust security due to their distributed nature. As big data grows exponentially, the gap between security needs and available protection mechanisms widens.\\
To address these challenges, we introduce the Dynamic Database Chaos Encryption (DaChE) Algorithm. DaChE leverages chaos theory to create a dynamic, self-protecting environment for sensitive data. Unlike traditional static encryption methods, DaChE employs chaotic behavior to continually transform data, ensuring robust protection while maintaining database performance. This approach offers security against a wide range of threats with minimal impact on system efficiency.\\
The key innovation of DaChE lies in its dynamic application of chaos theory, where data is in constant motion, protected by unpredictable yet deterministic behavior. This innovation provides a new paradigm in database security, offering a flexible and adaptive defense against sophisticated cyber threats.\\
This paper provides a detailed analysis of the DaChE algorithm, exploring its theoretical foundations in chaos theory and its practical implementation in modern database systems. We demonstrate how this novel approach achieves superior security while preserving essential database functionality. In doing so, we propose DaChE as a robust solution to the growing security challenges in today’s complex digital landscape.\\
We also discuss the potential application of chaos theory models, such as chaotic billiards, in understanding the dynamic security behaviors of DaChE and its resistance to multiple types of attacks.

\subsection{Types of Chaos}
In chaos theory, chaotic systems are classified based on their transition to chaos, the nature of their attractors, and their dynamical behavior, with systems categorized into levels depending on their response to predictions. Level 1 chaotic systems remain unaffected by predictions, while Level 2 chaotic systems adjust based on predicted outcomes, known as First and Second Order Chaos, respectively \cite{goertzel2013chaotic}. Level 3 chaotic systems, such as complex financial models using Type-3 fuzzy logic, are difficult to predict using traditional linear methods and are confirmed as chaotic through Lyapunov Exponents (LE)\footnote{The Lyapunov characteristic exponent measures the rate at which infinitesimally close trajectories diverge \cite{young2013mathematical}.} and attracting dimension tests. These systems exhibit complex interactions and display less extreme chaos with more balanced dynamics \cite{tian2022new}. \\
Chaos theory also identifies various chaotic behaviors or types, such as Lorenzian chaos, "Sandwich" chaos, and "Horseshoe" chaos which can be described by 3-dimensional state spaces of two simple non-linear differential equations \cite{rossler1976different}.\\
In addition, we can find Periodic Chaos, Deterministic Chaos, Bifurcation Chaos, Strange Attractor Chaos, Hyperchaos, Spatiotemporal Chaos, Noise-Induced Chaos, and Fractional Chaos, each with unique characteristics of unpredictability, sensitivity to initial conditions, and complex dynamics \cite{wernecke2019chaos}. These types of chaos appear in systems ranging from physics and engineering to economics and cybersecurity, with the motion of a billiard ball in a chaotic system, displaying deterministic chaos, serving as a classic example of how sensitive a system can be to initial conditions.\\
Sheela et al. presents types of chaotic functions and demonstrate properties of chaotic systems are more superior to pseudo random number generators \cite{sheela2017application}.\\
In this paper, chaos is implemented using a billiard stadium framework. This approach offers several benefits, such as the potential for implementation in a 2D chip, which will be discussed later. However, it is important to emphasize that the billiard model is just one way to implement chaos, and any DaCHE system can integrate with other chaotic systems that support the desired level of user security.\\
The motion of a billiard ball in a chaotic system serves as a classic example of deterministic chaos. The system is highly sensitive to initial conditions—small changes in the starting point can lead to vastly different outcomes. Similarly, DaChE's encryption dynamically responds to subtle shifts in user access patterns, ensuring that data remains secure and unpredictable, even in the face of advanced cyber threats.

\subsection{DaCHE algorithm Motivation}
The proliferation of NoSQL databases and big data has introduced unique security challenges that traditional protection mechanisms struggle to address. While these modern databases offer unprecedented scalability and flexibility, they often prioritize availability and partition tolerance over strict data consistency, as outlined in the CAP theorem\footnote{The CAP (Consistency, Availability, and Partition) theorem states that in a distributed system, it is possible to achieve only two out of the three essential attributes tolerance.} \cite{munoz2019cap}. This architectural trade-off, combined with the limitations of conventional security measures, creates significant vulnerabilities.\\
Traditional encryption techniques, including symmetric and asymmetric cryptography \cite{curtmola2006searchable}, face increasing challenges in protecting modern data architectures. Their primary limitation lies in their static nature - predetermined encryption keys and algorithms remain fixed over time, making them vulnerable to brute-force attacks and advancing computational capabilities. \\
In addition, storing data in a DaCHE-protected database can add an extra layer of protection against potential post-quantum cracking of traditional encryption methods and is aligned with the National Institute of Standards and Technology's (NIST) ongoing development of post-quantum cryptography standards.\\
Furthermore, the sheer volume and velocity of big data have strained the capabilities of conventional security measures. Conventional encryption techniques, while still valuable, struggle to keep pace with the evolving threat landscape, as attackers devise increasingly sophisticated methods to circumvent these defenses. The need for a more robust and adaptive approach to database security has never been more pressing.\\
As a result we can find in Table \ref{tab:database-security} the drawbacks of countermeasure approaches to common database attack vectors such as SQL injections, Phishing, Malware, Unpatched software, Cloud database configuration errors, Third-party vendors/service providers, Insider threats, Lack of encryption, Misconfigurations, Cross-site scripting, Man-in-the-middle attacks, Session hijacking, Weak Authentication, Privilege abuse, Exploiting unpatched services, and Insecure system architecture.
These threats are compounded by the limitations of traditional countermeasures.\\
SQL injections allow malicious code insertion into database queries, while phishing attacks deceive users into revealing sensitive information. Malware poses risks through malicious software designed to damage or infiltrate systems, often exploiting unpatched software running outdated versions with known vulnerabilities. Cloud database configuration errors and misconfigurations in general settings create significant vulnerabilities, while third-party vendors and service providers introduce additional security risks through external access points. \\
Insider threats from individuals with legitimate access, combined with lack of encryption leaving sensitive data exposed, present persistent challenges. Cross-site scripting enables injection of malicious scripts into trusted websites, while man-in-the-middle attacks intercept communications to steal or manipulate data. Session hijacking allows unauthorized capture of valid user sessions, particularly dangerous when combined with weak authentication methods that are easily compromised. Privilege abuse through misuse of legitimate access rights and exploitation of unpatched services further compound these issues, while insecure system architecture introduces fundamental design flaws that compromise security. \\
Traditional countermeasures, while necessary, each have significant limitations: firewalls, which monitor and control traffic based on predetermined rules, cannot protect against insider threats or inherent database vulnerabilities, becoming completely ineffective if compromised. Access Control Lists (ACLs) manage user permissions but face challenges with outdated static permissions and complex permission management across organizations. Encryption, while essential for data protection, doesn't prevent application layer vulnerabilities, creates performance overhead, and requires complex key management.\\
Input validation, critical for verifying user-supplied data, heavily depends on proper developer implementation and cannot prevent all attack types. Auditing provides crucial but only retrospective security insights through activity logging, requiring secure storage and continuous monitoring. Segmentation/isolation strategies, though effective for breach containment, still leave interconnected systems vulnerable and add significant management complexity. Authentication systems, while fundamental for access control, can be compromised to grant broad system access and may contain vulnerabilities themselves. Monitoring provides real-time threat observation but only detects issues after occurrence and relies heavily on rapid human response and expertise. Finally, patching systems address known vulnerabilities but leave systems exposed between updates and require careful change management.\\
These threats are compounded by the limitations of traditional countermeasures.
While Table \ref{tab:database-security} outlines the drawbacks of common countermeasures to these database attack vectors, it becomes evident that traditional defense mechanisms are often inadequate in addressing the full spectrum of complex and evolving threats that databases face.\\
In this context, the development of the DaChE Algorithm represents a timely and critical response to the security challenges posed by the contemporary data ecosystem. By harnessing the principles of chaos theory, this conceptual framework aims to redefine the boundaries of data protection, safeguarding the most valuable and sensitive information assets from the looming threat of database breaches.\\
The DaChE Algorithm addresses these challenges by integrating topology and chaos theory principles. This novel approach creates a dynamic, adaptive security framework that:
\begin{itemize}
    \item Exploits data's inherent structural properties
    \item Introduces unpredictable yet deterministic protection mechanisms
    \item Maintains high performance while ensuring robust security
    \item Adapts to evolving threats in real-time
\end{itemize}
The topological underpinnings of the DaChE algorithm enable the exploitation of data's inherent structural properties, allowing for the development of encryption techniques that are resilient to tampering and more resistant to cryptanalytic attacks. Furthermore, the chaotic dynamics inherent in the algorithm's design can introduce unpredictability and uncertainty, making it increasingly difficult for adversaries to gain a foothold in the system.\\
By harnessing the power of chaos theory, our proposed algorithm promises to redefine the boundaries of database security, delivering a more robust and adaptable solution that can keep pace with the ever-changing demands of the digital age.

\begin{table}
\begin{tabular}{p{6cm} p{6cm}}
\toprule
Solutions & Drawbacks \\
\midrule
Firewalls & Don't protect against insider threats or database vulnerabilities themselves \\
 & Compromised firewall provides no additional protection \\
\midrule
Access Control Lists (ACLs) & Static permissions can become outdated \\
 & Complex permissions are difficult to manage \\
\midrule
Encryption & Doesn't prevent vulnerabilities in database application layer \\
 & Performance overhead of encryption operations \\
 & Key management is critical \\
\midrule
Input Validation & Relies on developers implementing correctly \\
 & Doesn't prevent other types of attacks completely \\
\midrule
Auditing & Provides only after-the-fact visibility \\
 & Relies on logs being secured and monitored \\
\midrule
Segmentation/Isolation & Interconnected systems still have potential vulnerabilities \\
 & Complex to manage isolated systems and data flows \\
\midrule
Authentication & Compromise of credentials provides broad access \\
 & Vulnerabilities in the authentication system itself \\
\midrule
Monitoring & Detects threats after-the-fact \\
 & Reliant on rapid response and analyst abilities \\
\midrule
Patching & Assumes all vulnerabilities are patched systems remain vulnerable between patch cycles \\
 & Relies on change control processes \\
\bottomrule
\end{tabular}
\caption{Current Solutions and their Drawbacks in Database Security}
\label{tab:database-security}
\end{table}

\subsection{From Data in Rest to Motion}
In the domain of data management, conventional methods involve storing databases in specific locations like dedicated servers or cloud infrastructure. These databases strictly adhere to the ACID principles—Atomicity, Consistency, Isolation, and Durability—ensuring data integrity and transaction reliability \cite{lotfy2016middle}.\\
Relational algebra, a cornerstone of database management, provides a structured framework for defining and manipulating data within relational databases \cite{barr2006relational}. It comprises operations such as selection, projection, join, and set operations, facilitating complex data transformations and queries.\\
Contrasting the static nature of databases, protocols govern the transmission of data between systems. These protocols establish rules and standards for data exchange, including syntax (data structure and format), semantics (meaning of data), and timing (sequence of data exchange).\\
Data in motion go beyond traditional databases, focusing on the continuous, real-time transmission of digitally encoded information. Defined as a constant flow of data items arriving at a system over time, data streams are consumed in their arrival order without system control over timing or sequence.
This shift from static data to dynamic streams poses new challenges and opportunities in data management and processing. Specialized techniques and technologies are essential to handle the continuous and dynamic nature of streaming data effectively.

\subsection{Database in Motion Chaos Encryption (DaChE) Algorithm Advantages}
\label{DaChEAdvantages}
DaChE Algorithm provides key advantages including simulation capabilities, minimal overhead, energy-based fine-tuning, full Structured Query Language (SQL) \cite{halvorsen2016structured} functionality support, chaos-based encryption implementation, fast computation through on-the-fly processing with parallel computation alignment, highest security certification (EAL\footnote{An Evaluation Assurance Level (EAL) in a security service refers to the degree of confidence in its security. The EAL levels defined in the Common Criteria (ISO 15408) outline specific requirements for conducting an IT security audit \cite{kou2008definition}.} level), and complete support for ACID properties and relational algebra operations. In addition, as mentioned earlier, DaCHE can help protect against potential post-quantum cracking of traditional encryption methods.\\
The DaChE algorithm introduces a novel approach to securing data in motion while maintaining comprehensive database functionality. By leveraging chaos-based encryption with tunable energy parameters, it provides robust security without compromising performance. The algorithm's sophisticated ability to fine-tune data energy levels enables precise control over encryption strength, a critical feature particularly valuable since data spends the majority of its lifecycle at rest in databases. This energy-based approach allows organizations to adapt security levels based on specific requirements and threat landscapes.\\
DaChE distinguishes itself by maintaining full SQL functionality and ACID properties while supporting complex relational algebra operations. This ensures seamless integration with existing database systems and applications without sacrificing security. Its innovative on-the-fly computation approach, combined with inherent parallel processing capabilities, ensures minimal overhead and exceptional execution speed, making it ideal for high-performance environments. Moreover, increasing the initial energy through the velocity of the balls can lead to faster collisions, resulting in higher convergence of the system. This characteristic enables faster execution of SQL commands and database manipulation.\\
The algorithm's achievement of the highest EAL certification demonstrates its superior security capabilities and compliance with the most stringent security requirements, as there is a physical and mathematical proof and explanation for the security. \\
Through its energy-based fine-tuning mechanism, organizations can optimize the balance between security requirements and computational resources, allowing for dynamic adjustment of encryption strength based on data sensitivity and performance needs. This adaptability, combined with minimal computational overhead, makes DaChE an exceptional choice for organizations requiring both high security and efficient database operations.\\
The simulation capabilities of DaChE further enhance its utility, allowing organizations to test and optimize their security configurations before deployment. This feature, coupled with its support for parallel computation, ensures that the algorithm can scale effectively while maintaining its security guarantees and performance characteristics.

\section{The DaChE Algorithm: Redefining Database Security through Chaos}
The DaChE algorithm leverages the fundamental principles of chaos theory to provide a highly secure and decentralized approach to data management and encryption. Unlike traditional database systems, where data is stored in defined locations, DaChE employs a chaotic data distribution model where no one, including the system owner, knows the precise location of each individual data shard.\\
Furthermore, the timing of SQL command commitments is inherently unpredictable, as the system ensures that commands will be eventually committed, but the exact timing remains uncertain due to the chaotic nature of the underlying processes. This unique combination of unknown data locations and unpredictable timing introduces significant challenges for adversaries, making DaChE a highly robust and secure solution for protecting sensitive information in dynamic, distributed, and unpredictable environments.
\begin{algorithm}
\caption{Database in motion Chaos Encryption (DaChE) Algorithm}
\label{alg:dache}
\begin{algorithmic}[1]
\State User define: $N$ (No. of shards), $M$ (No. of balls), $P$ (No. of obstacles), Select Billiard Stadium Type
\State Shard the database: $D \rightarrow {d_1, d_2, ..., d_N}$ where $D = \bigcup_{i=1}^N d_i$
\State Make the data move in a chaotic way (Map) \Comment{See Algorithm 2 for details}
\While{SQL command initiation is detected}
\State Process all relevant shard data (option for "on fly" operations)
\State Wait for system convergence (i.e. collision of all the balls with their matched obstacles)
\State Collect the partial results and make the required calculations (Reduce)
\EndWhile
\State Print the total results
\State Go to Step 1
\end{algorithmic}
\end{algorithm}
As mentioned above, the DaChE algorithm introduces a breakthrough approach to securing database systems, leveraging the principles of chaos theory and data motion. At the core of this innovative strategy is the sharding of database data, where the information is divided into smaller, manageable fragments. These data shards are then set in motion, traversing the system in a random and chaotic manner, as described in Algorithm \ref{alg:dache}.\\
The key feature of the DaChE algorithm is its ability to process SQL commands in a distributed and dynamic fashion. Whenever a SQL command is initiated, the algorithm springs into action, processing the relevant shard data in a parallel manner, with the option for "on-the-fly" operations \footnote{The SQL command can be implemented on the shard data stored in the ball.}. This distributed processing approach ensures that the data is never static, making it inherently difficult for attackers to locate and access sensitive information.\\
The key to aligning the chaotic movement of the sharded data with the processing of SQL commands lies in the parallel and distributed nature of the DaChE algorithm's data processing. When a SQL command is initiated, the algorithm springs into action, leveraging the MapReduce paradigm \footnote{A MapReduce program consists of a map function that handles filtering and sorting tasks, and a Reduce function that performs an aggregation operation \cite{karloff2010model}.} to efficiently handle the relevant data shards.\\
The cyclical nature of the DaChE algorithm, as depicted in Algorithm \ref{alg:dache}, ensures that the data is continually sharded, set in motion, and processed in a chaotic manner. This perpetual state of flux creates a formidable barrier against potential attackers, who must contend with the ever-changing landscape of the database system.
The Map phase of the process involves the parallel processing of the individual data shards. As the sharded data moves in a chaotic manner throughout the system, the DaChE algorithm seamlessly integrates this motion with the on-the-fly processing of the data. Each data shard is independently processed, with the option to perform "on-the-fly" operations as the shard traverses the system.\\
This parallel processing of the shards allows the algorithm to keep pace with the chaotic movement of the data, ensuring that the relevant information is always accessible and ready for further computation. The distributed nature of the Map phase ensures that the processing is not limited by the constraints of a single, centralized system, but rather leverages the collective computing power of the entire database infrastructure.\\
Following the Map phase, the system enters the REDUCE phase, where the partial results from the individual data shards are collected and consolidated. The DaChE algorithm then performs the necessary calculations to generate the final, comprehensive result in response to the initial SQL command.
The REDUCE phase is implemented as the balls enter a phase of convergence, allowing the partial results from the individual shards to be collected and consolidated. The DaChE algorithm then performs the necessary calculations to generate the final result, which is subsequently displayed to the user.\\
By embracing the MapReduce approach and aligning it with the chaotic motion of sharded data, the DaChE algorithm represents a paradigm shift in database security. This method efficiently processes SQL commands by leveraging parallel, on-the-fly processing of data shards and the distributed nature of the system, allowing it to adapt seamlessly to the constantly shifting landscape of the database. Harnessing the power of chaos theory, the DaChE algorithm challenges the traditional static nature of data storage, offering a robust and resilient solution that safeguards critical information assets against evolving threats in modern database environments.

\subsection{DaChE's Data in Motion part}
The following algorithm describes the data in motion component of DaChE, which manages dynamic data exchanges between various obstacles (boxes) and data shards (balls) within the system. Each obstacle is responsible for handling encryption keys and processing data associated with the balls when they collide. The process ensures secure data handling and efficient transfer of results to a central Master\footnote{Master coordinates the tasks running the MapReduce job.}. The steps outlined in the algorithm cover the randomization of ball and obstacle locations, the assignment of keys, and the decryption and processing of data upon collisions.\\

\begin{algorithm}[H]
\caption{DaChE's Dynamic Data in Motion}
\label{alg:dache_data_motion}
\begin{enumerate}
\item For each Obstacle $(O_P)$ randomize location $(x_P,y_P)$
within the stadium boundaries and locate it on the Stadium
\item For each $d_i$ $(i=1$ to $N)$ (data shard) 
\begin{enumerate}
\item match a random ball $B_j$ $(j=1$ to $M)$
\item generate symmetric key and share it with random Obstacle $(O_j)$ $(j=1$ to $P)$ \Comment{Each Obstacle/ box can contain one or more encryption keys}
\end{enumerate}
\item When a ball collides with an obstacle:
\begin{enumerate}
\item If the symmetric key is the same (i.e., collided ball key = obstacle key):
\begin{enumerate}
\item The box (locally) decrypts the ball's data
\item Runs the relevant SQL statement (If it has not been processed on-the-fly)
\item Forward the result to the 'Master' (The Master accumulates the results from all the boxes)
\end{enumerate}
\end{enumerate}
\end{enumerate}
\end{algorithm}

\begin{figure}[ht]
\centering
\includegraphics[width=0.6\textwidth]{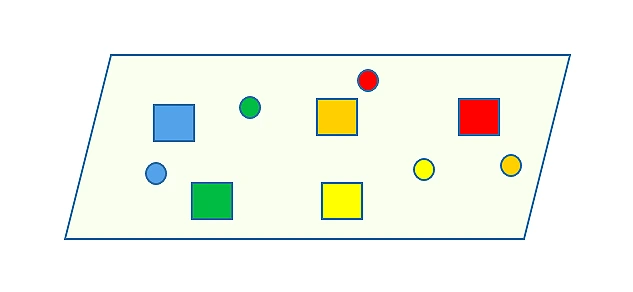}
\caption{A billiard system illustrating chaotic dynamics: each ball represents a shard of encrypted data along with an encrypted key, while the boxes, acting as obstacles, contain a list of encrypted keys (balls of the same color represent matching keys).}
\label{fig:billiard}
\end{figure}

As shown in Figure \ref{fig:billiard}, the dynamics of the billiard system exhibit chaotic behavior, where the location of the encrypted data shards is unknown at any given time.
As mentioned above, the DaChE algorithm disrupt the traditional approach to database security that challenges the traditional static nature of data storage as can be seen in Algorithm \ref{alg:dache_data_motion}. At the core of this strategy lies the principle of Dynamic Data Motion (DDM), where the data is sliced and encrypted into "balls" that are then set in motion throughout the system. The number of these balls is a configurable system parameter, allowing for fine-tuning and optimization.

There are several options for mimicking chaos motion. In this paper, we employ a billiard motion mimic. The billiard motion of the balls containing the sharded data moves chaotically in two dimensions. This facilitates the implementation, simulation of the algorithm, and its development on a chip.\\
The algorithm begins by choosing a stadium pattern that defines the movement and interaction of the balls within the database system. Alongside this, the algorithm randomly scatters "obstacles" (boxes/anchors) throughout the environment. For each ball, the algorithm randomly selects an obstacle and sends one of the ball's encryption keys to that obstacle, creating a dynamic association between the data and the security elements.\\
The critical moment occurs when a ball collides with an obstacle. The algorithm checks if the symmetric key held by the obstacle matches the key embedded within the ball. If a match is found, the obstacle locally decrypts the ball's data, runs the relevant SQL statement, and forwards the result to a central Master node. This Master node then accumulates the results from all the obstacles.\\
By leveraging the principles of chaos theory and data motion, the DaChE algorithm introduces a paradigm shift in database security. This novel approach makes it inherently difficult for attackers to locate and access sensitive data, as the data is constantly in flux, and the security elements are dynamically associated with the data through the encrypted balls and their randomly scattered obstacles. This unique combination of data dynamism and security chaos provides a robust defense against evolving threats, redefining the boundaries of data protection in the modern database landscape.

\section{Theoretical Foundations of Physical Billiard Chaos}
As mentioned earlier, our DaCHE algorithm splits the database into shards and moves them in a chaotic manner. One way to move the data shards is by mimicking the motion of billiard balls. A key advantage of this approach is that it achieves chaotic movement in two dimensions. In this section, we demonstrate how moving the data using billiard ball dynamics is chaotic.

\subsection{Chaos Theory: Fundamental Principles and Applications in Cryptography}
Chaos theory, a branch of mathematics studying dynamic systems profoundly impacted by initial conditions, highlights how even the smallest alteration in a chaotic system's input can yield vastly divergent outputs. These fundamental principles find significant application in cryptography. \\
Chaotic systems, marked by acute initial condition sensitivity, showcase the "butterfly effect," where minute starting point adjustments lead to significantly altered long-term behaviors. Despite deterministic unpredictability—embodying behavior guided by precise mathematical equations—forecasting the long-term trajectory of chaotic systems becomes progressively intricate over time.\\
The study of billiards has played a pivotal role in advancing chaos theory, providing a straightforward yet intricate model for studying nonlinear dynamics. Insights gained from analyzing billiard ball motion have been applied across scientific and engineering fields, including fluid mechanics, plasma physics, cryptography, and computer science. \\
In the context of DaCHE, we leverage these chaotic properties to ensure unpredictable data shard movement patterns, high sensitivity to initial positioning and efficient space coverage through ergodic trajectories.

\subsection{Chaos Complexity of Billiards}
The classic billiards table, with its smooth, frictionless surface\footnote{In practice, when designing a chip, there is a need to find a solution to the frictionless assumption.} and rigid bouncing walls, offers an excellent model for exploring chaos theory \cite{chernov2006chaotic}. Here, the motion of a billiard ball follows the principles of a Hamiltonian dynamical system\footnote{A Hamiltonian system is a dynamic system controlled by Hamilton's equations. This system characterizes the progression of a physical system \cite{mackay2020stability}.}.

\subsection{Sensitivity to Initial Conditions}
One of the defining characteristics of chaos is extreme sensitivity to initial conditions \cite{gradoni2023chaos}. In the billiard stadium, even minute changes in the initial position or velocity of the billiard ball can lead to vastly different trajectories over time. This sensitivity is captured by the LE, which quantifies the rate of divergence of nearby trajectories.

\subsection{Nonlinear Dynamics and Mixing}
The motion of the billiard ball within the stadium is governed by nonlinear dynamics, as the collisions with the curved and straight walls introduce complex interactions \cite{chernov2005billiards}. These nonlinearities, combined with the mixing properties of the stadium geometry, further contribute to the chaotic behavior observed.

\subsection{Billiard Two Dimensions (2D) Chaotic Behavior}
The dynamics of billiards can be classified based on the geometry of the billiard stadium \cite{gutkin2012billiard}: Circular billiard (integrable), Elliptical billiard (integrable), Sinai billiard (chaotic), Bunimovich stadium (chaotic), Rectangular billiard (integrable with rational angle ratios), and Polygonal billiard, Elliptical billiard, and Circular billiard.\\
The LE gives the average exponential rate of divergence of nearby trajectories.
We refer to the positive part of the LE as \( h \), because in the case we are dealing with, \( \lambda_+ \) is equal to \( h \), the Kolmogorov-Sinai entropy. \( h \) is defined as the sum of all the positive parts of the LE, minus \( E \), the escaping rate. In our system, there is no escaping, and the LE has four parts: one is positive, one is negative, and two are zeros $(E = 0)$.\\
As established in the literature, there is a formula involving the Kolmogorov-Sinai (KS) entropy in the case of the Sinai billiard (a square with a single disk in the middle), which is a function of the middle disk \cite{dellago1996lyapunov,garrido1997kolmogorov}. This formula also incorporates the average time between collisions with the disk, denoted as \( \langle \tau \rangle \):
\[
\langle \tau \rangle h = -2 \ln(R)
\]
where \( R \) represents the radius of the disk. We derived an expression for a generalized \( \langle \tau \rangle \), as a function of the number of disks and their radius:
\[
\langle \tau \rangle = \frac{1}{2 r n}
\]
where \( r \) is the radius of the disks and \( n \) is the number of disks.\\
Now, let us consider the case where we have a few disks located randomly. The KS entropy (or LE) increases as the number of convex surfaces grows \cite{papenbrock2000lyapunov}. 

\subsection{Positive Lyapunov Exponents (LE)}
Quantitatively, the chaotic nature of the billiard stadium can be demonstrated by the presence of positive LEs. For the billiard stadium, researchers have calculated the LEs and found them to be positive, confirming the system's chaotic behavior.\\
The combination of sensitivity to initial conditions, nonlinear dynamics, and mixing properties, all reflected in the positive LEs, collectively contribute to the complex and unpredictable motion of the billiard ball within the stadium geometry. These properties make billiard dynamics an ideal choice for the DaCHE algorithm's data movement strategy.

\subsection{Chaotic Billiard System with Obstacles}
We propose a chaotic billiard system where, in addition to the inherently chaotic stadium geometry, we introduce obstacles within the arena. These obstacles play a crucial role during the REDUCE phase of the process, where they collect results initiated by users. The placement of these obstacles further enhances the chaotic properties of the system, as they disrupt the continuous flow of trajectories, creating new regions in the phase space and making the system even more unpredictable.\\
In this configuration, the obstacles act like additional balls within the billiard arena, amplifying the chaotic behavior. Their presence disrupts the topological mixing properties of the system by breaking the continuous trajectory flow, leading to a more intricate interaction between regular and irregular motion. This increased complexity enhances the system's chaotic nature, where the sensitivity to initial conditions causes nearby trajectories to diverge rapidly, making long-term predictions practically impossible.\\
The combined effects of the stadium geometry and the strategically placed obstacles generate a system with unpredictable dynamics. The system exhibits both ordered and chaotic motion, a characteristic feature of chaotic systems. This duality of behavior offers new possibilities for applications in fields such as cryptography, where the inherent unpredictability can be used to bolster data security.\\
Now, let us consider the mathematical model for a particle moving in this chaotic billiard system. Imagine a billiard board with a disk (obstacle) of radius \(r\) randomly positioned inside. We aim to determine the time it will take for a particle, starting from a random location, to collide with this obstacle. The probability density function \(p(t)\) for the collision time \(t\) is given by the exponential distribution \cite{freitas2014convergence}:
\[
p(t) = \lambda e^{-\lambda t}
\]
where \(p(t)\) represents the probability density for the time until the first collision, and \(\lambda\) is the collision rate parameter. The time between collisions follows this memoryless exponential distribution, which is characteristic of random, independent events in a chaotic system.\\
To offer some intuition for this result, in a chaotic and ergodic system, we assume that the probability of colliding with the obstacle remains constant over time, denoted by \(p\). The Poisson distribution is a limiting form of the Binomial distribution that occurs when the number of trials, n, becomes very large and the probability of success, p, is small. The probability of the first collision occurring at the \(n\)-th step is given by \cite{dettmann2009survival}:
\[
(1-p)^{n-1} p
\]
For small \(p\), this simplifies to:
\[
\frac{p}{1-p} e^{n \ln(1-p)}
\]
Using approximations for small \(p\), we find the continuous limit:
\[
\frac{dP}{dt} = p e^{-pt} = \lambda e^{-\lambda t}
\]
Next, consider the variations of the Sinai and Bunimovich billiard boards by adjusting the proportions of the boards, area, and the particle’s velocity. This leads to the expression for the collision rate \(\lambda\) \cite{posch2000simulation}:
\[
\lambda \approx \frac{2r |v|}{A}
\]
where \(r\) is the radius of the obstacle, \(v\) is the velocity of the particle, and \(A\) is the area of the board. This formula suggests that, during an average time step, the particle "sees" a region of length \(\sqrt{A}\), but the obstacle occupies a fraction of this region, specifically \(2r\).\\
When more obstacles are added to the system, the LE increases, making the system more chaotic. However, the time for a particle to collide with a specific disk still follows the same exponential distribution as with a single obstacle. This suggests that the system's level of chaos does not influence the collision times directly.\\
Finally, in cryptographic applications, we consider a scenario with multiple obstacles scattered across the board, each associated with a corresponding particle. With random positions for the obstacles and particles, the time for all particles to collide with their respective disks can be modeled. Assuming all disks are of equal size and the particles do not interfere with each other, the probability that all particles collide with their corresponding disks by time \(t\) is given by:
\[
P(T_1, \dots, T_n \leq t) = \left[ \int_{0}^{t} \lambda e^{-\lambda t} \, dt \right]^n = \left( 1 - e^{-\lambda t} \right)^n
\]
The corresponding probability density is:
\[
\frac{dP}{dt} = n \lambda e^{-\lambda t} \left( 1 - e^{-\lambda t} \right)^{n-1}
\]
This expression highlights the relationship between the number of obstacles, the collision rate, and the probability of a successful collision, providing a foundation for understanding how this chaotic system could be used to enhance data security in cryptographic contexts.

\section{Evaluation of DaChE framework}
The evaluation of the DaChE algorithm spans three critical dimensions: Functionality, Time and Complexity, and Security Level. In terms of functionality, DaChE adeptly supports essential relational algebra operators, enabling seamless handling of a vast array of SQL commands crucial for data management tasks within existing database frameworks. In addition, DaCHE algorithm supports ACID properties and Relational Algebra, as mentioned in Section \ref{DaChEAdvantages}. \\
Evaluating its performance in terms of time and complexity involves assessing energy consumption, a pivotal metric shaping overall runtime and system intricacy, with potential for system operators to fine-tune this parameter to strike a balance 
between performance optimization and resource efficiency. 
The system user can control the initial velocity of the balls, thereby increasing their collisions. In this way, the system executes SQL commands faster.\\
Delving into security considerations, the robustness of DaChE can be gauged through the EAL framework, leveraging chaos theory principles to bolster its mathematical underpinnings and formal verification processes for enhanced system security. Furthermore, assessing the algorithm's "Chaotic Level" offers insights into its security efficacy by measuring the protection conferred by chaotic system attributes like sensitivity to initial conditions, deterministic unpredictability, and attractor dynamics—a higher Chaotic Level signaling greater resilience against potential breaches. By scrutinizing DaChE across these dimensions, encompassing functionality, time and complexity, and security level, researchers and practitioners can effectively gauge the algorithm's efficiency, effectiveness, and resilience, ensuring its viability for real-world deployment in secure and dynamic data management environments.\\
By leveraging the chaotic nature of the billiard stadium and strategically placing obstacles within the arena, user can create a highly unpredictable and resilient system. This enhanced chaotic behavior could have significant implications in fields such as cryptography, where the inherent unpredictability can be harnessed to improve data security and integrity.
Increasing the initial energy through the velocity of the balls as well as the number of obstacles can lead to faster collisions, resulting in higher convergence of the system. This characteristic enables faster execution of SQL commands and database manipulation.\\

\section{Discussion}
The performance characteristics of the DaChE algorithm are shaped by several key factors that determine its scalability, efficiency, and overall effectiveness. Some of the primary factors influencing scalability are the number of data shards, balls, and obstacles in the system, as well as the stadium's size and configuration. The interaction between these elements dictates how the system can handle increasing volumes of data while maintaining speed and reliability.\\
The algorithm is designed to minimize performance overhead by leveraging parallel processing of data shards, executing on-the-fly operations during ball movements, and efficiently managing key matching during collisions. Additionally, the system uses an optimized MapReduce implementation to ensure that data is processed in a distributed manner, further enhancing overall system performance. This parallelization is crucial for maintaining high throughput, especially as the number of balls, obstacles, and data shards grows.\\
In terms of security, DaChE offers robust protections through its continuous motion of data and the use of encryption at each step. Data confidentiality is ensured by encrypting the balls' data, which is only decrypted upon collision with a matching obstacle. This mechanism also protects the system against static analysis, as the unpredictable movement of data shards complicates any attempts to trace or analyze them. Moreover, the algorithm resists timing attacks due to the erratic collision patterns that do not follow predictable timing intervals.\\
The system also incorporates robust failure handling mechanisms, such as redundant data sharding and checkpoint-based recovery. In the event of an obstacle failure, keys can be redistributed automatically, ensuring that the system remains operational even in the face of partial failures. This approach contributes to the resilience of the DaChE algorithm, allowing it to gracefully degrade without compromising data integrity or availability.\\
Key management is another critical aspect of DaChE's design. The algorithm employs dynamic key generation for each data shard, ensuring that keys are fresh and unique. Secure key distribution is maintained through encrypted channels, and regular key rotation ensures that no single key is used for an extended period. This dynamic key management strategy minimizes the risk of key exposure and provides an added layer of security. \\
From a real-world deployment perspective, the implementation of DaChE requires careful attention to hardware and network considerations. The system’s performance depends on the optimization of hardware resources, as well as the management of network latency and bandwidth. Integrating DaChE with existing database systems requires a seamless approach, ensuring compatibility with the broader infrastructure. Regular monitoring and maintenance are essential to ensure that the system runs efficiently and securely over time.\\
While DaChE excels in performance and security, certain edge cases must be carefully managed. These include environments with extremely high-frequency transactions, systems with limited resources, or scenarios that demand deterministic response times and ultra-low latency. Addressing these challenges requires tuning system parameters, applying specialized optimization techniques, and leveraging hardware acceleration when necessary. Custom configurations for specific use cases can also help overcome some of these limitations, ensuring that DaChE remains adaptable and effective across a wide range of deployment scenarios.

\section{Conclusions}
The DaChE algorithm represents a significant shift in database security by integrating chaos theory with dynamic data motion. This innovative approach brings several important contributions to the field.
One of the core features of DaChE is its use of chaos-based security, which leverages the inherent properties of chaotic systems to protect database content. This approach not only ensures data security but also maintains efficient response times. The strong mathematical and physical foundation of chaos theory, along with formal security proofs, underpins the algorithm’s reliability and robustness.\\
In addition to its security mechanisms, DaChE provides security architects with valuable tools for proactive risk management. Through simulation capabilities and chaotic level assessment metrics, the algorithm enables the evaluation and optimization of security strategies, ensuring that potential risks are addressed before they can impact the system.
The algorithm is also highly flexible, supporting various levels of granularity and database partitioning. This adaptability allows organizations to tailor their security measures to meet specific requirements. DaChE can be deployed modularly across different database segments, offering scalable security enhancements and customizable protection levels to suit a wide range of operational needs.\\
Despite its robust security features, DaChE introduces minimal overhead, ensuring operational efficiency. This makes the algorithm practical for real-world deployment, where system performance cannot be sacrificed for security.
By seamlessly integrating encryption with chaos theory-based data motion, DaChE redefines data protection. The algorithm provides a comprehensive and adaptable solution for safeguarding critical information against evolving security threats, addressing both data-in-motion security and the implications of chaotic motion in modern database systems.

\bibliographystyle{IEEEtran}
\bibliography{ref.bib}

\end{document}